% Differentialoperator article   Clausthal talk
%\title
%Differential Operator Algebras
%on compact Riemann Surfaces${ }^1$
%\endtitle
%\author Martin Schlichenmaier \endauthor
%\address
%Department of Mathematics and Computer Science,
%University of Mannheim,
%D-68131 Mannheim, Germany
%\endaddress
%
%
% amstex 2.1
%
\input amstex
\documentstyle{amsppt}
%
%  To include encapsulated Postscript images
%
\input psbox
\let\fillinggrid=\relax
\loadbold
%
% local macros
%
%   Macros for Differentialoperatoralgebras
%
%
\font\ssf=cmss10
%  Blackboard-Bold
%
\def\R{\Bbb R }

\def\P{\Bbb P }

\def\Z{\Bbb Z }
\def\C{\Bbb C }

\def\Eins{\hbox{\ssf 1\kern-1.0pt{\vrule height6.5pt depth0pt}}}
%
% additional fonts
%
%\font\bigbf=cmbx10 scaled 1200
%\font\bigbff=cmbx10 scaled 1440
%\font\small=cmr8
%
%

\define\nl{\hfil\newline}
\define\ldot{\,.\,}
\redefine\l{\lambda}

\define\a{\alpha}
\redefine\d{\delta}

\redefine\b{\beta}
\redefine\t{\tau}
\redefine\i{{\,\bold i\,}}

\define\LMP{Lett\. Math\. Phys\. }
\def\CMP{Commun\. Math\. Phys\. }
\define\JMP{Jour\.  Math\. Phys\. }

\define\FA{Funktional Anal\. i\. Prilozhen\.}
\define\mapleft#1{\smash{\mathop{\longleftarrow}\limits^{#1}}}
\define\mapright#1{\smash{\mathop{\longrightarrow}\limits^{#1}}}
\define\mapdown#1{\Big\downarrow\rlap{
   $\vcenter{\hbox{$\scriptstyle#1$}}$}}
\define\mapup#1{\Big\uparrow\rlap{
   $\vcenter{\hbox{$\scriptstyle#1$}}$}}
\define\cint #1{\frac 1{2\pi\i}\int_{C^{#1}}}
\define\cintt{\frac 1{2\pi\i}\int_{C_{\tau}}}
\define\cinttt{\frac 1{24\pi\i}\int_{C_{\tau}}}
\define\im{\text{Im\kern1.0pt }}
\define\re{\text{Re\kern1.0pt }}
\define\res{\text{res}}
\define\RS{Riemann surface}
\define\RSs{Riemann surfaces}

\redefine\deg{\operatornamewithlimits{deg}}
\define\ord{\operatorname{ord}}
\define\fpz{\frac {d }{dz}}

\define\dzl{\,{dz}^\l}
\define\pfz#1{\frac {d#1}{dz}}
\define\pfx#1{\frac {d#1}{dX}}
\define\pft#1{\frac {d#1}{dt}}
\define\KNN {Krichever - Novikov }
\define\KN{{\Cal K\Cal N}}
\define\XA{X\setminus A}
\define\Fl{\Cal F^\lambda(A)}
\define\Fln #1{\Cal F_{#1}^\lambda(A)}
\define\Dln #1{\Cal D_{\lambda,#1}^1(A)}
\define\Dn #1{\Cal D_{#1}^1(A)}
\define\Fn#1{\Cal F^{#1}(A)}
\define\glb{\overline{gl}(\infty)}

\define\HE{\Cal H}
\define\HW{{{\Cal H}^\l}}

\define\K{{\Cal K\Cal N}}
\define\DA{{\Cal D}}

\define\DE{{\widehat{\Cal D}^1}}
\define\DO{{\Cal D^1}}
\define\KE{\widehat{\K}}

\magnification=1200
%
%
%    allgemeine Layoutparameter
%
\hoffset=0.4cm\voffset=0.4cm
\hsize=13.5cm\vsize=18.5cm
\baselineskip=13pt plus 0.2pt
\parskip=4pt
%
%\NoBlackBoxes
\TagsOnRight
%
%%%%%%%%%%%%%%%%%%%%%%%%%%%%%%%%%%%%%%%%%%%%%%%%%%%%%%%
%%%%%%%%%%%%%%%%%%%%%%%%%%%%%%%%%%%%%%%%%%%%%%%%%%%%%%
%    Numbering of references
%
\newcount\refCount
\def\newref#1 {\advance\refCount by 1
\expandafter\edef\csname#1\endcsname{\the\refCount}}

\newref ACKP % Arbarello-Deconcini-Kac-Procesi
\newref BREM % Bremner
\newref DICK % Dick
\newref EGA % EGA IV
\newref GUN % Gunning Riemann surfaces
\newref GUO % Chinese group
\newref HASCH % Hawley/Schiffer proj.connect
\newref JKL % Jaffe, Klimek und Lesniewski   Heisenberg
\newref KACRAI % Kac-Raina booklet
\newref KNFA % Krichever-Novikov Funkt.Anal.
\newref LI % Li (cocycle)
\newref RADUL % cocycle
\newref RDS  % Ruffing, Deck, Schlichenmaier (Tori)
\newref SADOV % Sadov CMPe
\newref SCHLL % Schlichenmaier  LMP articles
\newref SCHLTH % Schlichenmaier thesis
\newref SCHLDEG % Schlichenmaier (degenerations)
\newref SCHLC % Schlichenmaier (big article)
\newref SHEIN % Sheinman (Kac-Moody)
\hfill Mannheimer Manuskripte 164

\hfill hep-th/9311036

\hfill October 1993
\vskip 1cm
\topmatter
\title
Differential Operator Algebras
on compact Riemann Surfaces${ }^1$
\endtitle
\rightheadtext{Differential Operator Algebras}
\leftheadtext{Martin Schlichenmaier}
\author Martin Schlichenmaier \endauthor
\address
Department of Mathematics and Computer Science,
University of Mannheim,\nl
D-68131 Mannheim, Germany
\endaddress
\email
schlichenmaier\@math.uni-mannheim.de
\endemail
\abstract
This talk reviews results on the structure of
algebras consisting of meromorphic differential operators
which are holomorphic outside a finite set of points on  compact
Riemann surfaces.
For each partition into two disjoint subsets
of the set of points where poles are allowed,
 a grading of the algebra and of the modules
of $\l-$forms is introduced. With respect to this grading
the Lie structure of the algebra and of the modules are almost
graded ones. Central extensions and semi-infinite wedge representations
are studied.
If one considers only differential operators of degree 1 then
these algebras are  generalizations of the Virasoro algebra
in genus zero, resp\. of Krichever Novikov algebras in higher genus.
\endabstract
\endtopmatter
%
%
%    here the article starts
%
{\sl ${ }^1$ invited talk at the International Symposium on
Generalized Symmetries in Physics at the
Arnold-Sommerfeld-Institute, Clausthal, Germany, July 26 -- July 29, 1993}
\vskip 1cm
\head
1. Introduction
\endhead
It  is well-known that the Virasoro algebra
together with its representations
play a fundamental role in Conformal Field Theory (CFT).
Let me recall the realization of the Virasoro algebra which is suitable for
the generalization I am considering.
The meromorphic vector fields on $\P^1$ (everything is over the
complex numbers) which are holomorphic over $\P^1\setminus\{0,\infty\}$
can be given as
$p(z)\fpz$, with $p\in\C[z,z^{-1}]$ a Laurent polynomial.
The commutator
$$[\,p(z)\fpz,r(z)\fpz\,]= \big(p(z)\pfz r(z)-r(z)\pfz p(z)\big)
\fpz\tag 1$$
defines a Lie algebra structure.
This Lie algebra is called the Witt algebra. Obviously a basis is
given by
$$ l_n:=z^{n+1}\fpz,\quad n\in\Z\ ,$$
and one calculates immediately
$\  [\,l_n,l_m\,]=(m-n)\,l_{n+m}$.
The Witt algebra admits a universal central extension $V$ with
one-dimensional center
$$ 0\ \mapright {}\  \C\ \mapright {} \ V\ \mapright {}
\  W\ \mapright {}\ 0 \ .$$
The algebra $V$ is the Virasoro algebra. A basis is given by
lifts $L_n\  $ of $\ l_n\ $ and a central element $t$.
The well-known structure equations are
$$
[\,L_n,L_m\,]=(m-n)L_{n+m}+\frac 1{12}(m^3-m)\delta_{m,-n}\,t,
\qquad [t,V]=0\ .\tag 2
$$
It is important to keep in mind that the Virasoro algebra
is not only this algebra but this algebra together
with the standard grading induced by
$\ \deg(L_n)=n$ and $\deg (t)=0 $.
{}From (2) we see immediately that $V$ is a graded Lie algebra.
Such a grading is essential
if one wants to  construct highest
weight representations (e.g\. by using semi-infinite wedge
forms). In CFT one is exactly looking for such
representations.
I do not want to go into the applications to CFT here
but let me just say that in string theory the Virasoro
algebra (or better the Witt algebra) describes the
situation of a world sheet of genus zero
with one incoming  string (at $P=\{z=0\}$) and one outgoing  string
(at $Q=\{z=\infty\}$) (see Fig.~1).
%
%
%   Figure 1 and Figure 2 (Virasoro and Krichever-Novikov situation)
%
\centinsert{
\hbox{
\pscaption{\psannotate
{\psboxto(4cm;0cm){dcfig1.eps}}{\fillinggrid
\at(4\pscm;3\pscm){$P$}
\at(9.5\pscm;3\pscm){$Q$}
}}
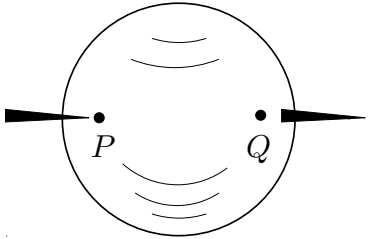
{Fig.~1: Virasoro situation ($g=0,
\  N = 2$)}}
\hskip 1cm
\hbox{
\pscaption{\psannotate
{\psboxto(6cm;0cm){dcfig2.eps}}{\fillinggrid
\at(4\pscm;3\pscm){$P$}
\at(20.5\pscm;2.8\pscm){$Q$}
}}
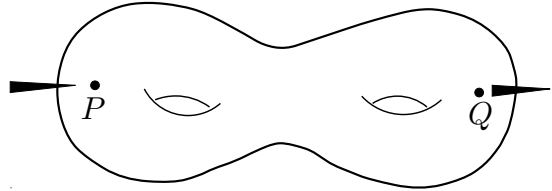
{Fig.~2: An example of a Krichever - Novikov situation}}
}   %end of centinsert
{}From this point of view it is quite natural to generalize the situation
to \RSs\ of arbitrary genus and an arbitrary (but fixed) number of
incoming strings and  an arbitrary
(again fixed) number of outgoing string.
For one incoming and one outgoing string this was done by
Krichever and Novikov (1987/88) \cite\KNFA\
(see Fig.~2) and for arbitrary numbers
 by myself (1989--) \cite\SCHLL,\cite\SCHLTH
\ (see Fig.~3).
%
%     Fig.3 (generalized situation)
%
\centinsert{
\pscaption{\psannotate
{\psboxto(9cm;0cm){dcfig3.eps}}{\fillinggrid
\at(5.5\pscm;3\pscm){$P_1$}
\at(5.5\pscm;6\pscm){$P_2$}
\at(20.5\pscm;5\pscm){$Q_1$}
}}
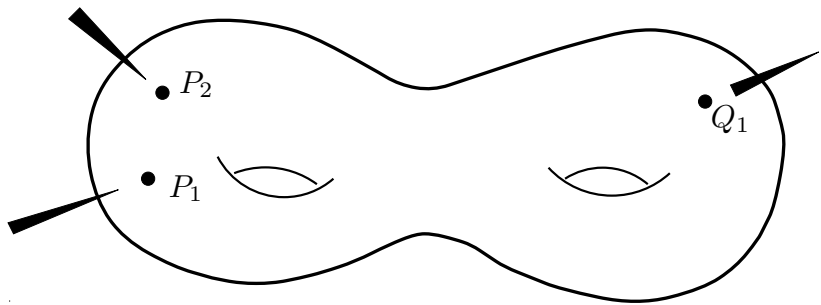
{Fig.~3: An example of a generalized situation
($N =3$, 2 in-points, 1 out-point) }
}   %end of centinsert
I will consider the more general
situation of  differential operators of  arbitrary degree. The degree
one case corresponds to the vector field case.
In this talk I will review some of the results obtained.
For more information I have to refer to \cite\SCHLTH\  and \cite\SCHLC.

Let me add that the many-point situation was also studied by
Dick \cite\DICK, by  Guo, Na, Shen,
Wang, Yu, Wu, Chang (see for example \cite\GUO), and by Bremner
(3 points and genus 0) \cite\BREM\  but without introducing a grading.
Only Sadov \cite\SADOV\  did some work
which is related to my approach.
\head
2. The set-up and the involved algebras
\endhead
For the following let $X$ be a \RS\ (or a  nonsingular algebraic curve
over
an arbitrary algebraically closed field $k$ with char $k=0$),
$A$ a finite set of points which is divided into two disjoint
subsets $I$ and $O$,
$\ A=I\cup O,\  \#I=k\ge 1,\  \#O=l\ge 1$,\nl $N=k+l$.
The elements of $I$ are the ``in-points'', the elements of
$O$ are the ``out-points''.
Let $\rho $ be a meromorphic differential
holomorphic outside $A$ with exact pole order
1
at the points of $A$, (given) positive residues at $I$,
(given) negative residues
at $O$ (of course, obeying $\sum_P\res_P(\rho)=0$), and purely
imaginary periods.
Fix a point $Q\in \XA$. Then the function
$$u(P)=\re \int_Q^P\rho\tag 3$$
 is a well-defined harmonic function.
The level lines
$$C_\t=\{P\in\XA\mid u(P)=\t\}, \quad \tau\in\R\tag 4$$
define a fibering of $\XA$ (see Fig.~4).
%
%
%     Fig.4 (level lines)
%
\centinsert{
\pscaption{\psannotate
{\psboxto(9cm;0cm){dcfig4.eps}}{\fillinggrid
\at(3.4\pscm;5\pscm){$\boldsymbol\tau\boldsymbol\ll \boldkey0$}
\at(17.6\pscm;4.4\pscm){$\boldsymbol\tau\boldsymbol\gg\boldkey 0$}
}}
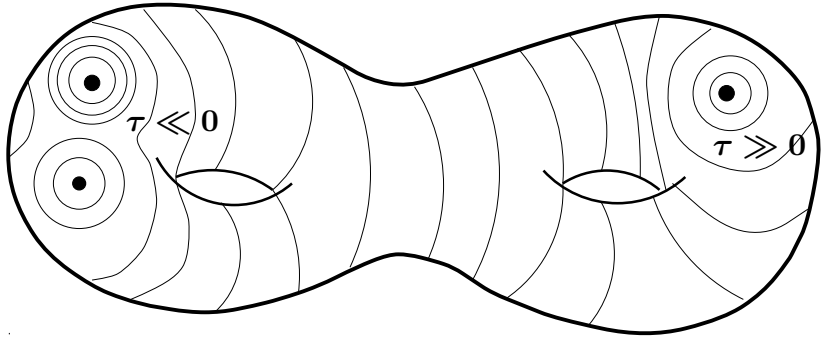
{Fig.~4: Fibering by level lines}
}   %end of centinsert
Every level line separates the in- from
the out-points.
For $\tau\ll 0$ the level line $C_\tau$ is a disjoint union
of deformed circles around the points in $I$.
For $\tau\gg 0$ it is
a disjoint union
of deformed circles around the points in $O$.
In the interpretation of string theory this $\tau$ might be interpreted
as proper time of the string on the world sheet.

Let $K$ be the canonical bundle, i.e\. the bundle whose
local sections are the local holomorphic differentials. For every
$\l\in\Z$ we consider the bundle $K^\l:=K^{\otimes\l}$,
the
bundle with local sections the forms of
weight $\l$.
After fixing a square root of the canonical bundle (a so-called
theta characteristic) everything below can even be done for
$\l\in \frac 12\Z$.
For simplicity we consider here only the case of integer $\l$.
We denote
by $\Fl$ the vector space of global meromorphic sections of $K^\l$ which are
holomorphic on $\XA$.
Special cases are  the differentials ($\l=1$),  the functions ($\l=0$),
and  the vector fields ($\l=-1$).
For the vector fields we also use the special
notation $\KN(A)$.

The functions (i.e\. the elements in $\Fn 0$)
 operate by multiplication on $\Fl$.
The vector fields (i.e\. the elements in $\KN(A)$)
operate by taking the Lie derivative on $\Fl$.
In local coordinates
the Lie derivative can be described as
$$
L_e(g)_|=(e(z)\fpz)\ldot (g(z)\dzl)=
\left( e(z)\pfz g(z)+\l\, g(z)\pfz e(z)\right)\dzl \ .\tag 5
$$
Here and in the following I will use the same symbol for the
section of the bundle and its local representing function.
By (5) $\KN(A)$ becomes a Lie algebra and the vector spaces
$\Fl$ become Lie modules over $\KN(A)$ (i.e.
$ [L_e,L_f]=L_{[e,f]}$).
The algebra $\KN(A)$ I call a generalized Krichever Novikov algebra.

For an arbitrary associative algebra $R$ we obtain
always a Lie algebra $\ LR\ $ with  same underlying vector
space and taking the commutator
$[a,b]=a\cdot b-b\cdot a$ as Lie product.
Obviously, $L\Fn 0$ is an abelian Lie algebra. We take the
semi-direct product
$$\DO(A)=L\Fn 0\times \KN(A)$$
to obtain the Lie algebra of differential operators of
degree $\le 1$ (which obey again the above regularity conditions).
Its structure is given by
$$
[\,(g,e),(h,f)\,]:=(\,L_eh-L_fg\,,\,[e,f]\,),\tag 6
$$
and we have the exact sequence of Lie algebras
$$
0\ \mapright {}\ L\Fn 0 \ \mapright {}\
\DO(A)\ \mapright {}\ \KN(A)\ \mapright{} \ 0\ .\tag 7$$
Note that $\KN(A)$ is the subalgebra of differential operators
of degree 1. The spaces $\Fl$  now become Lie modules over
$\DO(A)$.

If we want differential operators of arbitrary
degree we have to do some universal constructions.
I use the following constructions:
$$
\DA(A)=U\DO(A)/J,\quad\text{resp.}\quad
\DA_\l(A)=U\DO(A)/J_\l,
$$
where
$U\DO(A)$ is the universal enveloping algebra
of $\DO(A)$ (with multiplication $\odot$ and unit $\Eins$), and
$J$ resp\. $J_\l$ are the following
two-sided ideals
$$\gather
J:=(\;a\odot b-a\cdot b,\ \Eins-1\mid a,b\in\Fn 0\;),\\
J_\l:=(\;a\odot b-a\cdot b,\ \Eins-1
,\ a\odot e -a\cdot e +\l L_e(a)
\mid a,b\in\Fn 0, e\in \KN(A)\; ).
\endgather$$
All $\Fl$ are modules over $\DA(A)$, but only $\Fl$ is a module over
$\DA_\l(A)$.
It is easy to check that every element $D$ of $\DA(A)$ resp\.
of $\DA_\l(A)$ operates as differential operator on $\Fl$.
I call the elements of $\DA(A)$ the coherent differential
operators and the elements of
$\DA_\l(A)$ the differential operators on $\Fl$.
Because $\XA$ is affine every
algebraic differential operator can be represented by such
an element $D$ \cite\EGA.
\head
3. The grading
\endhead
It is possible to introduce a grading in $\Fl$ in such a way
that the homogeneous subspaces $\Fln n$ of degree $n$ are
finite-dimensional and that the Lie module structure is
an almost graded one. Of course, this is a crucial step. Nevertheless
I want to skip the details here because they can be found in
\cite\SCHLL. Essentially the grading is given by the zero order
at the points in $I$.
To give  an example: Let $k=l$, $\l\ne 0,1$, and
$\  I=\{P_1,P_2,\ldots,P_k\}\ $,
$O=\{Q_1,Q_2,\ldots,Q_k\}\ $
be points in generic positions.
Then there is for every $n\in\Z$ and every $p=1,\ldots,k$ up to
multiplication with a scalar a  unique element
$f_{n,p}^\l\in\Fl$ with
$$
\aligned
\ord_{P_i}(f_{n,p}^\l)&=(n+1-\l)-\d_{i,p},\qquad i=1,\ldots,k,\\
\ord_{Q_i}(f_{n,p}^\l)&=-(n+1-\l),\qquad \qquad i=1,\ldots,k-1,\\
\ord_{Q_k}(f_{n,p}^\l)&=-(n+1-\l)+(2\l-1)(g-1)\ .
\endaligned\tag 8
$$
This can be shown
either  by using Riemann-Roch type arguments
or by explicit constructions.
After fixing a coordinate at the points $P_i$ the scalar constant
can be fixed.
The  elements $f_{n,p}^\l$ for $n\in\Z$ and
$p=1,\ldots,k$ are a basis of $\Fl$. The homogeneous elements are
defined to be the elements of the space
$$\Fln n:=\langle\; f_{n,p}^\l\mid p=1,\ldots, k\;\rangle,
\quad\text{resp.}\quad
\Dn n:=\langle\; f_{n,p}^0,f_{n,p}^{-1}\mid p=1,\ldots, k\;\rangle
\ .$$
The basis elements obey the important duality relation
(after a suitable fixing of the scalar)
$$\cintt f_{n,p}^\l\cdot f_{m,r}^{1-\l}=\d_{n,-m}\cdot
\d_{p,r},\tag 9$$
where $C_\tau$ is any non-singular level line.
In fact, the grading depends on the numbering of the
points $Q_i$ but the induced filtration depends only
on the partition $A=I\cup O$.
Another essentially different partition induces a non-equivalent
filtration.
Note that the duality (9) is true in all
  cases not only in the example discussed
above.

An almost graded structure means
that the homogeneous subspaces are finite-dimens\-ion\-al and that there
are constants $K$ and $L$ not depending on $n$ and $m$ such that
$$[\,\Dn n,\Dn m\,]\subseteq \bigoplus_{h=n+m+K}^{n+m+L}\Dn h,\qquad
\Dn n\;.\;\Fln m\subseteq \bigoplus_{h=n+m+K}^{n+m+L}\Fln h\ .
\tag 10$$
In our situation $K=0$ and there exist explicit formulas for $L$
which depend on the genus $g$ and  on $N=k+l$.
In the Virasoro situation ($g=0$, 2 points) $L=0$ and everything
reduces to the well-known graded situation.
Here I like to stress the fact, that the conditions (10)
are necessary to construct semi-infinite wedge representations
(see Section~5).
\head
4. Central extensions
\endhead
As it is well-known it is not possible to obtain representations
of the Witt algebra with certain properties one is interested in
(irreducible, unitary highest weight representations).
Such representations exist only for the central extension of
the Witt algebra, the Virasoro algebra.
Hence, it is necessary to study central extensions also for the above
algebras. Central extensions can be given by 2-cocycles
(with respect to Lie algebra cohomology).
Such defining cocycles are known for the Virasoro situation.
Unfortunately  these cocycles are not defined invariantly.
Only after adding some suitable counterterms, we obtain
objects which make sense on arbitrary Riemann surfaces.

({\bf 1}) Let  $R$ be a holomorphic
projective connection
\footnote{
see the appendix for the definition}
, then
$$\chi(e,f)=\cinttt \left(
\frac 12(e''f'-e'f'')-R\cdot(e'f-ef')\right) dz
\tag 11$$
for $e,f\in\KN(A)\subseteq\DO(A)$,
defines  a nontrivial 2-cocycle of $\KN(A)$. By pull-back via
(7) we obtain a cocycle for
$\DO(A)$.
Recall,  we represent $e$ by $e(z)\fpz$ and the prime means
derivative of $e(z)$ with respect to $z$.
This cocycle is a generalization to the many-point situation
of the cocycle
introduced by Krichever and Novikov \cite\KNFA.

({\bf 2}) For the abelian Lie algebra
$L\Fn 0$ a non-trivial 2-cocycle is defined by
$$\gamma(g,h)=\cintt g\,dh\tag 12$$
for $g,h\in\Fn 0$. The centrally extended Lie algebra is the
generalized Heisenberg algebra.
For applications see for example \cite\JKL.
It is easy to show that $\gamma$ can be extended
to a cocycle on $\DO(A)$ via
$\gamma((g,e),(h,f)):=\gamma(g,h)$.

({\bf 3}) There is another cocycle
of $\DO(A)$ which  connects $L\Fn 0$ with $\KN(A)$.
Let  $T$ be an affine meromorphic connection
\footnote{
see the appendix for the definition}
with at most one pole at a point of $O$ then
$$\beta(e,g)=-\beta(g,e)=
\cintt(eg''+T\cdot eg')dz\tag 13
$$
for $e=(0,e)$ and $g=(g,0)\in \DO$
defines a 2-cocycle by obvious extension.

In any of the cases above the choice of another connection
does not change the cohomology class of the  cocycle.
Hence, the equivalence
class of the central extension does not depend on the chosen
connection.
Obviously, the cocycle is independent on the value of $\tau$
used to fix the integration curve $C_\tau$.
Of course, the above expressions would also define
a cocycle if we integrate along any
other closed curve on $\XA$.
By considering only integration along $C_\tau$ we obtain
cocycles which are local in the sense of Krichever and Novikov
with respect to the grading
of the algebra $\DO(A)$ induced by the partition of
$A$. A cocycle is called local if there are constants $M$ and $H$ such that
for every homogeneous $e,f\in\DO(A)$
$$\psi(e,f)=0\qquad\text{ if }
\deg(e)+deg(f)>H,
\quad\text{ or }
\deg(e)+deg(f)<
M.\tag 14$$
In our situation we have $H=0$, and $M$ depends on the genus and the
number of points in $A$.
Why is locality important? It allows us to extend the almost grading to
the centrally extended algebra by defining the central element $t$
to have degree zero.
Recall that the structure equations of the central extension
$$0\ \mapright {}\ \C\ \mapright  {}\  \DE(A)\
\mapright {r}\ \DO(A)\ \mapright {}\ 0\tag 15$$
defined by the cocycle $\psi$
is given by
$$[\Phi(e),\Phi(h)]=\Phi([\,e,h\,])+\psi(e,h)\cdot t,\tag 16$$
where $\Phi$ is a linear splitting map of $r$ (resp\. a lift
of the basis elements in $\DO(A)$) and $t$ is the image of $1$ in
$\DE(A)$.

\remark{Remarks}
{\bf 1.} For $g=0,N=2$ everything reduces to the well-known case.
In particular, we obtain again the 3 linearly independent cocycles
for $\DO(\{0,\infty\})$ studied by
Arbarello-DeConcini-Kac-Procesi \cite{\ACKP}.
They showed that $H^2(\DO(\{0,\infty\}))$ is 3-dimensional. Hence
every one-dimensional central extension is defined
up to equivalence by a certain
linear combination of the above given ones.

{\bf 2.} For the general situation this is not true anymore.
Different partitions of the points will yield different level lines
and in general non-equivalent central extensions. But the new cocycle
is not local anymore with respect to the first partition.
Hence,
\proclaim{Conjecture}
(a) Every local cocycle of $\DO(A)$ is cohomologous to a linear
combination of the above 3 cocycles (11), (12), (13).
\nl
(b) Every local cocycle of $\KN(A)$ is cohomologous
to the  cocycle (11).
\endproclaim

{\bf 3.}
If we consider $\DA(A)$ or $\DA_\l(A)$ as Lie algebras  then
$\DO(A)$ is a Lie subalgebra. But it is not clear (and in fact
not true) whether any of the above 3 cocycles can be extended to
the whole of $\DA_\l(A)$.
In the Virasoro situation  Radul introduced
for $\DA_0(A)$ a ``canonical cocycle'' \cite\RADUL\
(which of course is an
extension of a  certain linear combination of the above three
cocycles).
It was shown by Li \cite\LI\  that this is in fact the only one
which could be extended.
Using semi-infinite wedge representations I was able to show that
there is a central extension
$\widehat{\DA}_\l(A)$ (as Lie algebra !) of
$\DA_\l(A)$. The defining cocycle $\psi_\l$ restricted to $\DO(A)$
is local. If we accept the above conjecture than it has to
be
cohomologous to a certain  linear combination of the above cocycles.
In this case it can be calculated as
$$
[\psi_\l]=[\chi]+\frac {2\l-1}{2c_\l}[\beta]+\frac {-1}{c_\l}[\gamma],
\tag 17$$
where $\ c_\l:=-2(6\l^2-6\l+1)\ $ denotes the famous expression
from Mumford's formula and $[..]$ the cohomology class.

{\bf 4.}
It is possible to define higher genus analogues of untwisted
Kac-Moody algebras. For this let $\frak g$ be a Lie algebra with
$<.,.>$ an invariant, symmetric, bilinear form on $\frak g$
(i.e\. $<[x,y],z>=<x,[y,z]>$).
$\widehat{\frak g}=(\frak g\otimes\Fn 0)\oplus \C\, c$ is now
a Lie algebra if we define
$$
[x\otimes f,y\otimes g]=[x,y]\otimes f\cdot g -
<x,y>\gamma(f,g)\cdot c,\qquad [c,\widehat{\frak g}]=0
\tag 18
$$
where $\gamma $ is the cocycle introduced in (12).
For $g=0,\ A=\{0,\infty\}$ they coincide exactly with the untwisted
affine Lie algebras.
For $N=2$ the same construction has been given by
Krichever and Novikov \cite\KNFA\  and if additionally  $g=1$
they were examined in detail by Sheinman
\cite \SHEIN.
\endremark
\head
5. Further results
\endhead
Here space does not permit to explain the semi-infinite
wedge representations in detail. Roughly speaking, the semi-infinite
wedge spaces $\HE^\l(A)$ are ``restricted $\frac {\infty}2$ external
forms'' constructed from the elements of $\Fl$.
The elements of $\HE^\l(A)$  are called semi-infinite forms of
weight $\l$.
To extend the action of
$\DA_\l(A)$ (resp\. $\KN(A)$, or $\DO(A)$)
 on $\Fl$ to $\HW(A)$ one has to
``regularize the action'' and hence has to allow
for central extensions of these Lie algebras.
The regularization is done by embedding the algebras via their action
on $\Fl$ into $\glb$, the algebra of both-sided infinite matrices with
only finitely many diagonals.
Here the almost graded structure is crucial.
For the algebra $\glb$ the regularization and central extension is
well-known \cite\KACRAI.
Pulling back the action and the cocycle
 gives a regularized action for a centrally
extended algebra of $\DA_\l(A)$ (resp\. of $\KN(A)$, or of $\DO(A)$).
The pulled back cocycle is local. Assuming the conjecture to be true
it can be written in the form (17).

If we consider only the algebra $\KN(A)$ and if we require a
coherent action of $\KE(A)$ on every  $\HW(A)$ we obtain that
the central element operates as
$\ (-2)(6\l^2-6\l+1)\cdot identity$.

In this manner one obtains highest weight, resp\. Verma module
representations of the centrally extended algebras.

Let me just give a brief summary what else has been done.
It is possible to define an action of $\Fl$ and $\Fn {1-\l}$ on
$\HW(A)$ by wedging, resp\. contracting the semi-infinite forms.
By this one obtains a Clifford algebra structure (or a $b-c$ system
in physicist's language). There are interesting compatibility relations
of the $\DE(A)$ action and the $b-c$ action.
In addition there is a natural pairing between right
semi-infinite forms of weight $\l$ and left
semi-infinite forms of weight $1-\l$ induced by the pairing
(9).

In the  case of the torus ($g=1$) explicit calculations have been
done in \cite\RDS\  for the 2, 3 and 4 point case.
In these cases also degenerations of the
Riemann surface have been studied \cite\SCHLDEG.
Here it is advantageous to use the language of Algebraic Geometry.
It is interesting to note that starting from a two-point situation
one is forced to study many-point situations on Riemann surfaces of
lower genus.
\head
Appendix: Affine and projective connections
\endhead
Let $\ (U_\a,z_\a)_{\a\in J}\ $ be a covering of the Riemann surface
by holomorphic coordinates, with transition functions
$z_\b=f_{\b\a}(z_\a)=h(z_\a)$.
A system of local (holomorphic, meromorphic) functions
$\ R=(R_\a(z_\a))\ $ resp\. $\ T=(T_\a(z_\a))\ $
is called a (holomorphic, meromorphic) projective (resp\. affine)
connection if it transforms as
$$
R_\b(z_b)\cdot (h')^2=R(z_\a)+S(h),\qquad\text{with}\quad
S(h)=\frac {h'''}{h'}-\frac 32\left(\frac {h''}{h'}\right)^2\tag 19$$
the Schwartzian derivative, resp\.
$$
T_\b(z_b)\cdot h'=T(z_\a)+\frac {h''}{h'}\ .\tag 20$$
Here ${}'$ denotes differentiation with respect to
the coordinate $z_\a$.
Note that the difference of two affine (projective) connections
is always a usual (quadratic) differential.
\proclaim{Lemma}
(a) There is always a holomorphic projective connection.
\nl
(b) Given a point $P$ on $X$ there is always a meromorphic
 affine connection which is holomorphic
outside $P$ and has there at most a pole of order 1.
\endproclaim
\noindent
For a proof of (a) see for example \cite{\HASCH} or \cite{\GUN}.
A proof of (a) and (b) can also be found in \cite{\SCHLTH}.
%
%   References
%

%
\enddocument
\autojoin
\bye